\documentclass[a4paper]{JACoW-GSI}
\usepackage{graphicx}
\usepackage{url}
\newcommand{\be}{\begin{equation}}
\newcommand{\ee}{\end{equation}}
\newcommand{\bea}{\begin{eqnarray}}
\newcommand{\eea}{\end{eqnarray}}
\newcommand{\nn}{\nonumber}

\newcommand{\ket}[1]{\left| #1 \right\rangle}
\newcommand{\spur}[1]{\not\! #1 \,}

\begin{document}
\title{New Charm Spectroscopy: Insights from Theory}

\author[1]{F. De Fazio\thanks{fulvia.defazio@ba.infn.it}}
\affil[1]{Istituto Nazionale di Fisica Nucleare, Sezione di Bari
\\Via Orabona 4\\I-70124 Bari, Italy}

\maketitle \abstract We discuss several new observations of mesons
with open  charm. In particular, we consider $D_{sJ}(2317)$ and
$D_{sJ}(2460)$ and compare their  isospin violating decays into
$D^{(*)}_s \pi$ to the radiative decays analysed using  light-cone
QCD sum rules. The results support the interpretation of these two
mesons as ordinary $c{\bar s}$ states.

In the case of  $D_{sJ}(2860)$ and $D_{sJ}(2710)$ we compute the
strong decays in the heavy quark limit. Comparison of the results
with   recent  measurements of the  BaBar Collaboration leads to
identify $D_{sJ}(2710)$ with the first radial excitation of
$D_s^*$, while the identification is still uncertain in the case
of $D_{sJ}(2860)$.

\endabstract

\section{Introduction}
Starting from 2003, charm spectroscopy has entered a new era, due
to a series of intriguing observations both in the open  and in
the hidden charm sectors \cite{reviews}. Some of the newly
observed states can be  easily  classified  within the quark model
scheme, some others still wait for a proper identification. Here
we focus  on mesons with charm and strangeness, in particular on
the narrow states $D_{sJ}(2317)$ and $D_{sJ}(2460)$,  observed in
2003, and  on the states $D_{sJ}(2860)$ and $D_{sJ}(2710)$,
discovered in 2006 and in 2007, respectively. As a preliminary
step, we introduce a suitable classification of mesons with a
single heavy quark which can be derived in the infinite heavy
quark mass limit, exploiting the symmetries of QCD in such a
limit.

\section{Hadrons with a Single Heavy Quark}
In the infinite heavy quark mass limit, $m_Q \to \infty$, the QCD
Lagrangian is invariant under heavy quark spin and flavour
rotations, and an effective theory can be built, known as Heavy
Quark Effective Theory (HQET) \cite{HQET}.

Let us consider hadrons with a single heavy quark $Q$. When $m_Q
\to \infty$, $Q$ acts as a static colour source for the light
degrees of freedom ({\it ldf}) of the hadron. In the case of
mesons, considered here, {\it ldf} consist of the light antiquark
$\bar q$ and  gluons. In particular, the heavy quark spin $s_Q$ is
no more coupled to the {\it ldf} total angular momentum $s_\ell$,
given by ${\vec s}_\ell={\vec s}_q+{\vec \ell}$, where $s_q$ is
the light antiquark spin and $\ell$  its orbital angular momentum
with respect to $Q$. Therefore,  a heavy hadron can be labelled
not only according to its total spin ${\vec J}={\vec s}_Q+{\vec
s}_\ell$, but also to the value of $s_\ell$. An important
consequence is that states which differ only for the orientation
of $s_Q$ with respect to $s_\ell$ are expected to be degenerate,
and this allows to collect heavy mesons in doublets, the members
of which have the same value of $s_\ell$ and correspond to the two
possible orientations of $s_Q$ with respect to $s_\ell$. Finite
heavy quark mass corrections remove the degeneracy between the
members of a  doublet and induce a mixing between states with the
same spin-parity $J^P$ belonging to different doublets.

Let us consider the lowest doublets that can be built according to
this classification. For $\ell=0$ one has a doublet of states
$(P,P^*)$ with $J^P_{s_\ell}=(0^-,1^-)_{1/2}$ (we  refer to this
as to the {\it fundamental} doublet), while two doublets
correspond to $\ell=1$: $(P^*_0,P^\prime_1)$ with
$J^P_{s_\ell}=(0^+,1^+)_{1/2}$ and $(P_1,P_2^*)$ with
$J^P_{s_\ell}=(1^+,2^+)_{3/2}$. For the purposes of this paper, we
need to introduce also the two doublets corresponding to $\ell=2$:
$J^P_{s_\ell}=(1^-,2^-)_{3/2}$  and
$J^P_{s_\ell}=(2^-,3^-)_{5/2}$. For each doublet, one can consider
a tower of  states corresponding to the radial excitations.

In the heavy quark limit one can  predict whether these states are
narrow or broad. For example, strong decays of the states
belonging to the $J^P_{s_\ell}=(1^+,2^+)_{3/2}$ doublet to the
fundamental doublet with the emission of a light pseudoscalar
meson occur in $d$-wave. Since the rate for this process is
proportional to $|\vec p|^5$ (in general, to $|\vec p|^{2\ell+1}$,
$p$ being the light pseudoscalar momentum and $l$ the angular
momentum transferred in the decay), these states are expected to
be narrow. On the contrary, states belonging to the
$J^P_{s_\ell}=(0^+,1^+)_{1/2}$ doublet decay in $s$-wave, hence
they should be broad.

Experimental data collected up to 2003 show that heavy mesons fit
very well in this scheme, as can be argued looking at the
experimental values of masses and widths: the degeneracy condition
is better fulfilled by beauty mesons than by charmed ones, as
should be,  the $b$ quark being approximately three times heavier
than  charm. As for the  widths, in Table \ref{cubar} we collect
masses and widths of the $c{\bar u}$ states identified as the
members of the $J^P_{s_\ell}=(0^+,1^+)_{1/2}$ and
$J^P_{s_\ell}=(1^+,2^+)_{3/2}$ doublets  \cite{PDG}. From this
table one can see that the members of the
$J^P_{s_\ell}=(1^+,2^+)_{3/2}$ doublet are indeed narrow, while
the widths of the analogous states belonging to the
$J^P_{s_\ell}=(0^+,1^+)_{1/2}$ doublet are much broader.
\begin{table*}[tb]
\centering \caption{Masses and widths of $c{\bar u}$ states
belonging to the  $J^P_{s_\ell}=(0^+,1^+)_{1/2}$ and
$J^P_{s_\ell}=(1^+,2^+)_{3/2}$ doublets.} {\begin{tabular}{|c |c
|c |c |c| } \hline $s_\ell^P$ & $J^P$ & state & M  (MeV) &
$\Gamma$ (MeV) \\ \hline
 &
$0^+$ & $D_0$ & 2400
& $283\pm 24\pm 34$\\
 ${1 \over 2}^+$ &&&& \\
 & $1^+$ & $D_{1}^\prime$ & 2430
& $384 \pm^{107}_{75} \pm 74$ \\ \hline
 &
$1^+$ & $D_1$ & 2420
& $20.4 \pm 1.7$ \\
 ${3 \over 2}^+$ &&&& \\
 & $2^+$ & $D_{2}^*$ & 2460
& $43 \pm 4$ \\ \hline
\end{tabular}} \label{cubar}
\end{table*}

In the case of  mesons with charm and strangeness, known states
are those composing the fundamental doublet: $D_s(1968)$ and
$D_s^*(2112)$ and the two mesons which can be assigned to the
$J^P_{s_\ell}=(1^+,2^+)_{3/2}$ doublet: $D_{s1}(2536)$, whose
width is  $<2.3$ MeV, and $D_{s2}^*(2573)$ with measured width:
$\Gamma(D_{s2}^*)=20 \pm 5$ MeV \cite{PDG}. In the following, we
analyse the other  mesons with charm and strangeness recently
observed.

\section{$D_{sJ}(2317)$ and $D_{sJ}(2460)$}\label{dsjprimi}

In April 2003 the BaBar Collaboration reported the observation of
a narrow peak  in the $D_s\pi^0$ invariant mass distribution
  with mass close to $2.32$ GeV
and width consistent with the experimental resolution
\cite{BaBar2317}. The resonance, named $D_{sJ}(2317)$, was
observed in both the $\phi\pi^+$ and $\overline{K}^{*0}K^+$ decay
modes of  $D_s^+$. The peak was also found by reconstructing $D_s$
through $D_s \to K^+K^-\pi^+ \pi^0$. No evidence for
$D_{sJ}(2317)\rightarrow D_s\gamma, D^*_s \gamma$ and
$D_s\gamma\gamma$ was  found. The observation was  confirmed by
Belle~\cite{Belle continuo}, CLEO  ~\cite{CLEO} and Focus
Collaboration \cite{FOCUS2317}.

The  decay $D_{sJ}(2317) \to D_s\pi^0$ implies for $D_{sJ}(2317)$
natural spin-parity.  The helicity angle distribution of
$D_s\pi^0$ obtained by BaBar is consistent with the spin 0
assignment,  even though it does not rule out other possibilities;
the absence of a peak in the $D_s\gamma$ final state supports the
spin-parity assignment  $J^P=0^+$. The measured mass is below the
$DK$ threshold  ${\rm M}_{D^+K^0} = 2.36$ GeV.

Together with the $D_{sJ}(2317)$, CLEO Collaboration reported the
observation of
 a narrow resonance, $D_{sJ}(2460)$,  in the $D_s^*\pi^0$ system~\cite{CLEO},
 with mass close to 2.46~GeV
and width consistent  with the experimental resolution. Later on,
also  radiative decays of $D_{sJ}(2460)$ have been detected, with
measured branching fractions: $BR(D_{sJ}(2460) \to  D_s
\gamma)=(18 \pm 4)\, 10^{-2}$ and $BR(D_{sJ}(2460) \to
D_{sJ}(2317) \gamma)=(3.7 \pm^{5.0}_{2.4})\, 10^{-2}$ , while the
upper limit $BR(D_{sJ}(2460) \to  D_s^* \gamma)<8\%$ \cite{PDG}
was put. Angular analyses suggest the assignment $J=1$. The mass
of $D_{sJ}(2460)$ is below the $D^*K$ threshold ${\rm
M}_{D^{*+}K^0}=2.51$ GeV.

Being two states with $J^P=(0^+,1^+)$ their natural interpretation
would be as the components of the doublet with $s_\ell^P={1 \over
2}^+$. However, this interpretation raises several questions. The
first one stems from the comparison with potential model
predictions of the masses, which correspond to  larger values,
 above the threshold allowing isospin
conserving decays ($DK$ and $D^*K$ in the two cases). The second
one is that the members of the $s_\ell^P={1 \over 2}^+$ doublet
are expected to be broad, while the observed mesons are narrow.
Many interpretations have been provided since the original
discovery of these states \cite{reviews}. However there are
arguments to support the interpretation of $D_{sJ}(2317)$ and
$D_{sJ}(2460)$ as ordinary $c{\bar s}$ states,
 their narrowness being due to the  low mass forbidding
isospin conserving decays.

\begin{figure*}[tb]
\centering
\begin{tabular}{cc}
\includegraphics*[width=70mm]{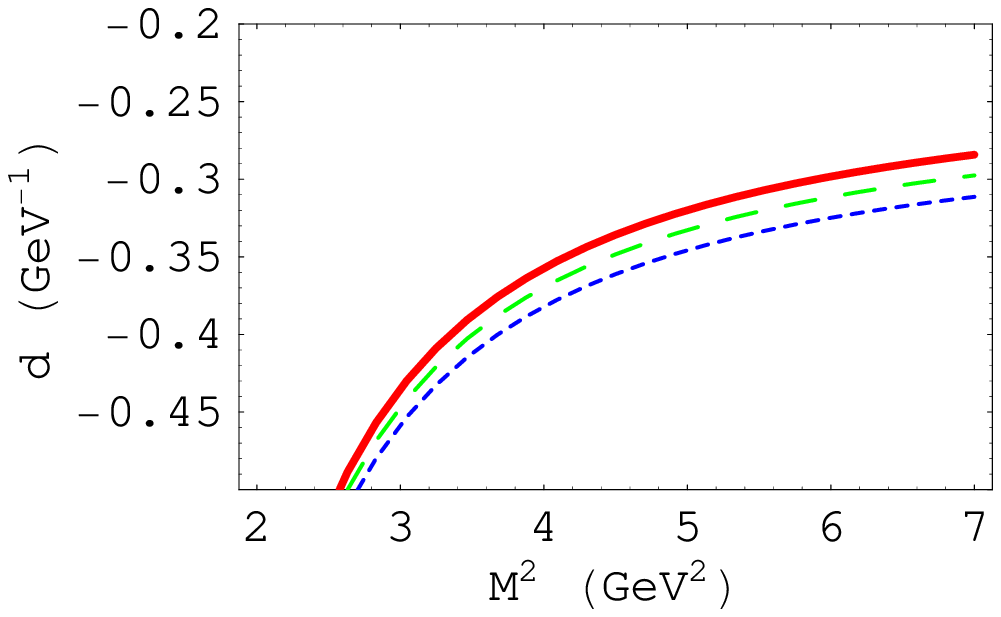}&\includegraphics*[width=70mm]{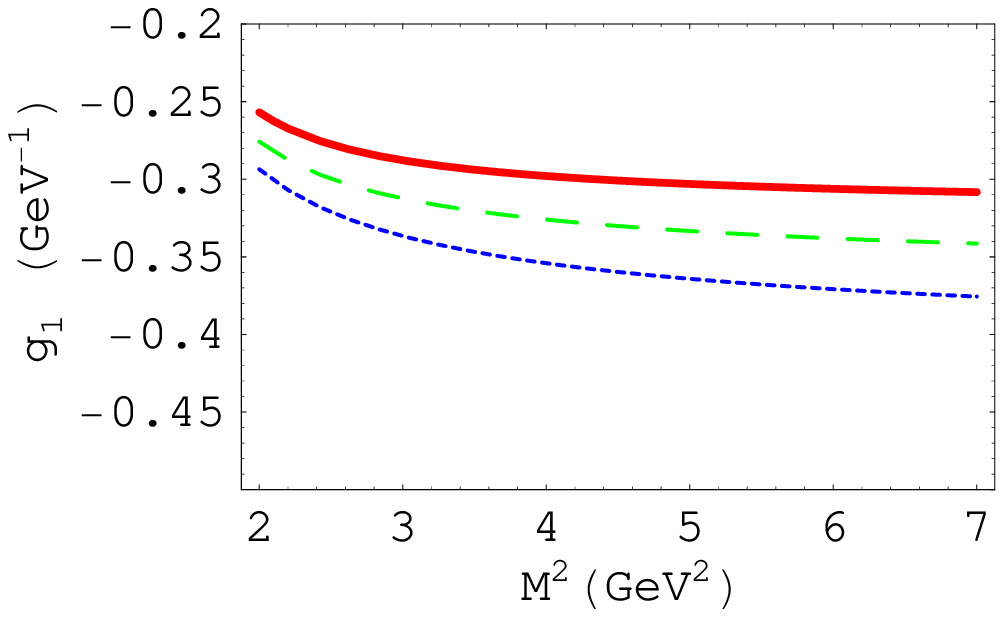}\\
\includegraphics*[width=70mm]{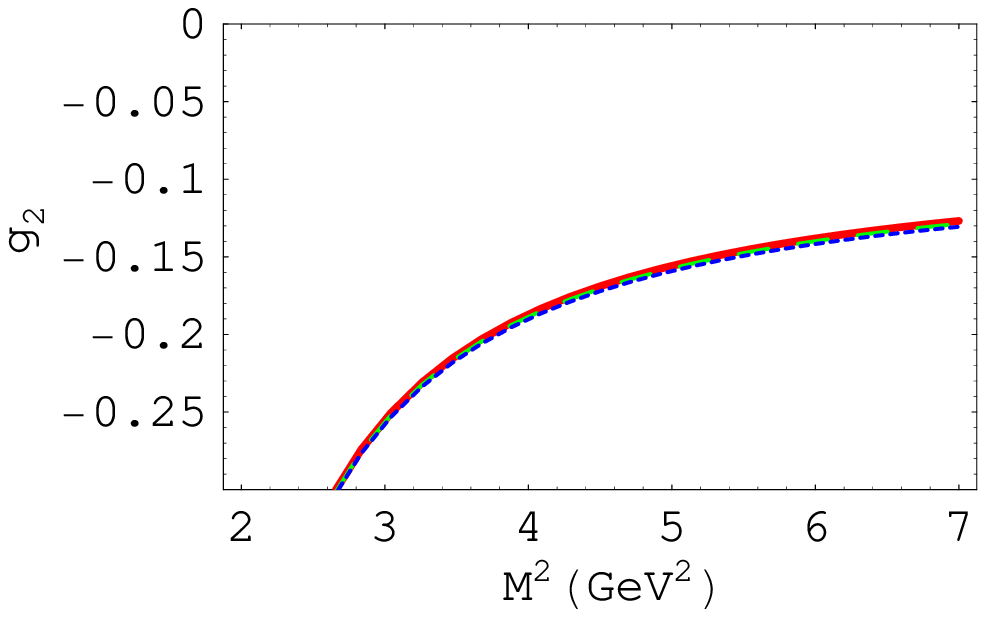}& \includegraphics*[width=70mm]{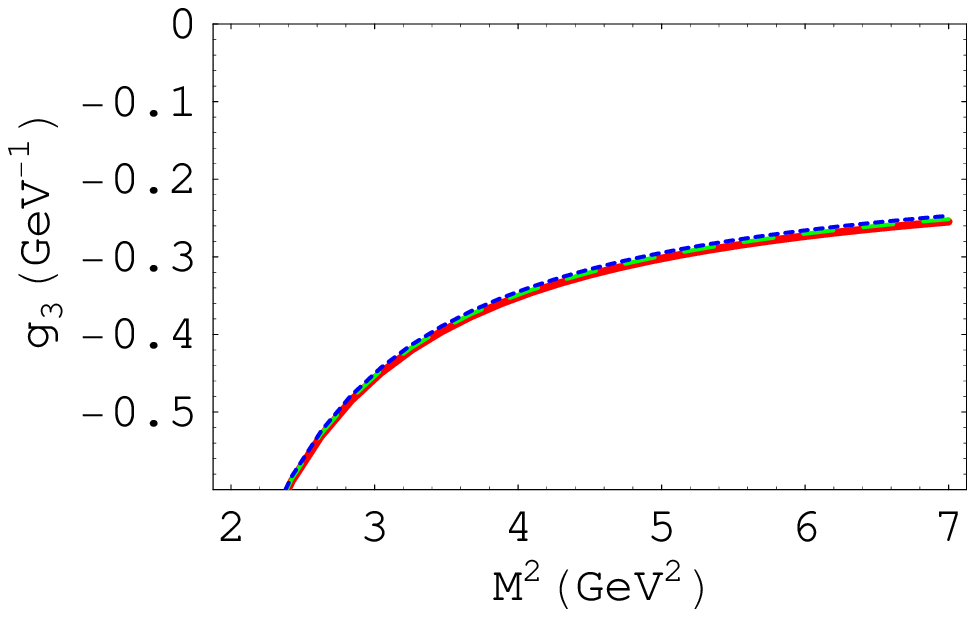}\\
\end{tabular}
\caption{Light-cone sum rule results for the hadronic parameters
governing radiative decays of $D_{sJ}(2460)$ and $D_{sJ}(2317)$;
$M^2$ is the Borel parameter.} \label{fig:results}
\end{figure*}

An example of such arguments is based on  the analysis of
radiative transitions, that probe the structure of hadrons
\cite{Godfrey:2003kg,Colangelo:2003vg}.
 Identifying $D_{sJ}(2317)$
 with $D_{s0}^*$ and $D_{sJ}(2460)$ with $D_{s1}^\prime$, the decay
 amplitudes  governing the $D_{s0}^* \to D_s^* \gamma$ and
 $D_{s1}^\prime \to D_s^{(*)} \gamma, \,  D_{s0}^* \gamma$ transitions:
\bea &&\langle \gamma(q,\lambda) D_s^*(p,\lambda^\prime)|
D_{s0}^*(p+q)\rangle \nonumber \\&&= e \, d  \left[ (\varepsilon^*
\cdot \tilde \eta^*)(p\cdot q)-(\varepsilon^* \cdot p)
(\tilde \eta^* \cdot q) \right]   \nn  \\
\nn \\
 &&\langle \gamma(q,\lambda) D_s(p)|
D_{s1}^\prime(p+q,\lambda^{\prime\prime})\rangle\nonumber \\ &&= e
\,g_1  \left[ (\varepsilon^* \cdot  \eta)(p \cdot
q)-(\varepsilon^* \cdot p)(\eta \cdot q) \right]
 \label{ampl-dsj}  \\
\nn \\
&&\langle \gamma(q,\lambda) D^*_s(p,\lambda^\prime)|
D_{s1}^\prime(p+q,\lambda^{\prime \prime})\rangle
\nonumber \\ &&= i \, e \, g_2 \,
\varepsilon_{\alpha \beta \sigma \tau}  \eta^\alpha \tilde \eta^{*\beta} \varepsilon^{*\sigma}
 q^\tau
 \nn \\
\nn \\
&&\langle \gamma(q,\lambda) D_{s0}^* (p)|
D^\prime_{s1}(p+q,\lambda^{\prime \prime})\rangle  \nonumber \\
&&= i \, e \, g_3  \, \varepsilon_{\alpha \beta \sigma \tau}
\varepsilon^{*\alpha} \eta^\beta    p^\sigma q^\tau \nn
 \eea
 involve the the hadronic parameters
 $d, g_1, g_2$ and $g_3$ ($\varepsilon(\lambda)$ is  the photon
 polarization vector and $\tilde \eta(\lambda^\prime)$,
$\eta(\lambda^{\prime\prime} )$ the $D_s^*$  and $D_{s1}^\prime$
polarization vectors). Such parameters can be computed by
light-cone sum rules  \cite{Colangelo:2005hv}. Considering the
correlation functions \cite{altri,Colangelo:2000dp} \be F(p,q)=i
\int d^4x \; e^{i p \cdot x} \langle \gamma(q,\lambda) |
T[J^\dagger_A(x) J_B(0)] |0\rangle \label{eq:corr-Ds0Ds*gamma} \ee
of quark-antiquark currents  $J_{A,B}$ having the same quantum
number of the decaying and of the produced charmed mesons,  and an
external photon state of momentum $q$ and helicity $\lambda$,
%
\begin{figure}[htb]
\centering
\includegraphics*[width=65mm]{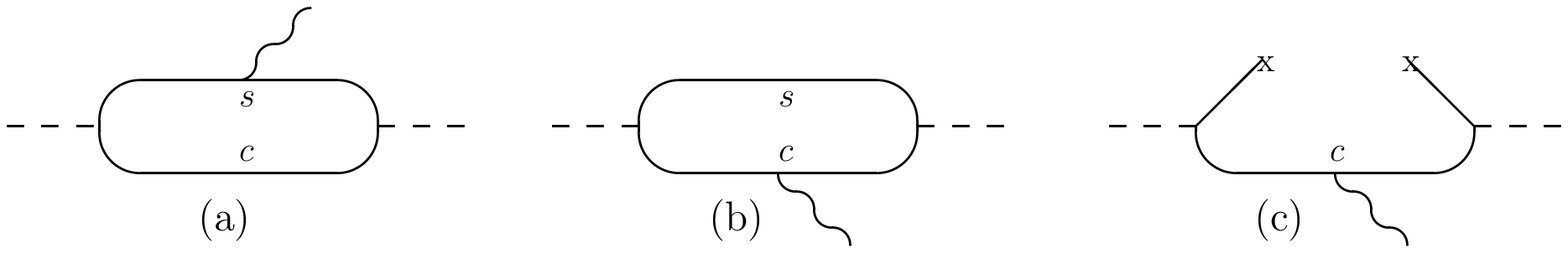}\\\includegraphics*[width=65mm]{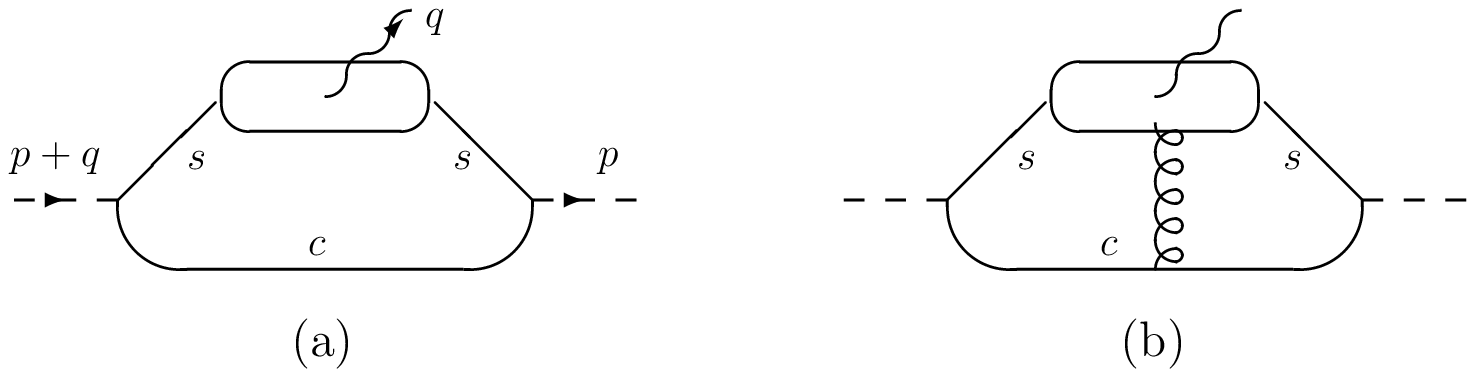}\\
\caption{Leading contributions to the correlation functions eq.
(\ref{eq:corr-Ds0Ds*gamma})
 expanded on the light-cone:
 perturbative photon emission by the strange  and  charm  quark ((a,b) in the first line)
 and  two- and three-particle photon distribution amplitudes (second line);
 (c)  corresponds to the strange quark condensate contribution.}
\label{fig:light-cone}
\end{figure}
%
\noindent and expanding on the light-cone, it is possible to
express $F$ in terms of the perturbative  photon coupling to the
strange and charm quarks,
 together with  the contributions of the photon emission from the soft $s$ quark,
expressed as photon matrix elements of increasing twist
\cite{Ball:2002ps}, see fig.\ref{fig:light-cone} . The hadronic
representation of the correlation function involves the
contribution of the lowest-lying resonances,  the current-vacuum
matrix elements of which are computed by the same method
\cite{Colangelo:1995ph},
 and a continuum of states  treated invoking  global quark-hadron duality.
The final step of the method consists in applying to both the
representations of the correlation function a
  Borel transformation, which improves in several respects the sum rule while
   introducing an external parameter $M^2$. The hadronic quantities should be
 independent of it, so that the final results are found requiring stability against variations
  of $M^2$ (fig. \ref{fig:results}).

\begin{table*}[tb]
\centering \caption{ Radiative decay widths (in keV) of
$D_{sJ}(2317)$ and $D_{sJ}(2460)$ obtained by Light-Cone sum rules
(LCSR),    Vector Meson Dominance (VMD) and constituent quark
model (QM).} {\begin{tabular}{|c |c |c |c |c| c|} \hline Initial
state & Final state & LCSR  \cite{Colangelo:2005hv}&VMD
\cite{Colangelo:2003vg} & QM \cite{Godfrey:2003kg}  & QM
\cite{Bardeen} \\ \hline
$D_{sJ}(2317)$&$D_{s}^{\ast }\gamma$&  4-6     & 0.85& 1.9 &  1.74\\
$D_{sJ}(2460)$   &$D_{s}\gamma$             & 19-29 & 3.3&  6.2 & 5.08 \\
                              &$D_{s}^*\gamma$          & 0.6-1.1 & 1.5&  5.5 & 4.66 \\
                               &\,\,\,\,\,$D_{sJ}(2317)\gamma$\,\,\,\,\,& 0.5-0.8  &  \
                               \hfill --- \hfill\ &0.012 & 2.74\\ \hline
\end{tabular}} \label{predictions}
\end{table*}

In Table \ref{predictions} the light-cone QCD sum rule results are
collected together with the results of other methods
\cite{Colangelo:2003vg,Godfrey:2003kg,Bardeen}. Looking at this
Table one can see that the rate of $D_{sJ}(2460) \to D_{s}\gamma$
is the largest one among the radiative $D_{sJ}(2460)$ rates, and
this is confirmed by experiment \cite{PDG}.
Quantitative understanding of  the  experimental data for both
hadronic and radiative decays requires a precise knowledge of the
widths of the isospin violating transitions $D_{s0}^*\to D_s
\pi^0$ and $D^\prime_{s1}\to D_s^* \pi^0$. In the description of
these transitions based on the mechanism of $\eta-\pi^0$ mixing
\cite{Godfrey:2003kg,Colangelo:2003vg} the accurate determination
of the strong   $D_{s0}^* D_s \eta$ and $D^\prime_{s1} D^*_s \eta$
couplings for finite heavy quark mass and including $SU(3)$
corrections is required.
These results suggest
 the identification of
$D_{sJ}(2317)$ and $D_{sJ}(2460)$ as the two members of the
$J^P_{s_\ell}=(0^+,1^+)_{1/2}$ doublet. Together with
$D_{s1}(2536)$ and $D_{s2}^*(2573)$, these two states fill the
four p-wave levels. In principle, the two $J^P=1^+$ mesons
$D_{s1}(2536)$ and $D_{sJ}(2460)$ could be a mixing of the
$s_\ell^P={1 \over 2}^+$ and $s_\ell^P={3 \over 2}^+$ states,
allowed at ${\cal O}(1/m_Q)$, an issue  discussed in the last
section.

\section{$D_{sJ}(2860)$ and $D_{sJ}(2710)$}

In 2006,  BaBar observed  another $c{\bar s}$ meson,
$D_{sJ}(2860)$, decaying to $D^0 K^+$ and $D^+ K_S$, with  mass
and width \cite{Aubert:2006mh}: \bea M(D_{sJ}(2860))
&=& 2856.6 \pm 1.5 \pm 5.0 \,\,  {\rm MeV}\nn \\
\Gamma(D_{sJ}(2860)) &=& 47 \pm 7 \pm 10 \,\,{\rm MeV} \,\, . \eea
 Shortly after,
  analysing the $D^0 K^+$  invariant mass distribution in
 $B^+ \to {\bar D}^0 D^0 K^+$  Belle Collaboration
  established the presence of a $J^P=1^-$
resonance, $D_{sJ}(2710)$, with \cite{Brodzicka:2007aa}:
\bea
M(D_{sJ}(2710)) &=& 2708\pm 9 ^{+11}_{-10} \,\, {\rm MeV} \nn \\
\Gamma(D_{sJ}(2710))&=& 108 \pm23 ^{+36}_{-31}\,\,
 {\rm MeV} \,\,. \eea

 To classify $D_{sJ}(2860)$ and $D_{sJ}(2710)$, one can  analyse the
strong decays, comparing the predictions which follow from
different quantum number assignments. This can be done using an
effective Lagrangian approach which exploits the symmetries that
Quantum Chromodynamics (QCD) exhibits in specific limits. One
 is chiral $SU(N_f)_L \times SU(N_f)_R$ symmetry holding in
the limit of $N_f$ massless quarks. This symmetry is spontaneously
broken to $SU(N_f)_V$ and light pseudoscalar mesons are identified
as Goldstone bosons acquiring mass when explicit symmetry breaking
mass terms are considered. An effective theory, chiral
perturbation theory, can be built as an expansion in the light
quark masses and momenta \cite{Gasser:1984gg}. The other one is
the heavy quark spin-flavour symmetry for $m_Q \to \infty$.

Interactions of heavy mesons with light ones can be described by
an effective Lagrangian displaying both
 heavy quark and  chiral symmetry.
 The Lagrangian was first formulated in the
 case of light pseudoscalars \cite{hqet_chir}, and  extended to
 include
  light vector mesons \cite{casalbuoni}.

In the heavy quark limit,  the doublets defined in the previous
Sections are described by effective fields: $H_a$ for
$s_\ell^P={1\over2}^-$ ($a=u,d,s$ is a light flavour index), $S_a$
and $T_a$ for $s_\ell^P={1\over2}^+$ and $s_\ell^P={3\over2}^+$,
respectively; $X_a$ and $X^\prime_a$ for
  $s_\ell^P={3\over2}^-$ and $s_\ell^P={5\over2}^-$, respectively:
 \bea
&& \hskip -0.6cm H_a  = \frac{1+{\rlap{v}/}}{2}[P_{a\mu}^*\gamma^\mu-P_a\gamma_5]  \label{neg} \nn  \\
&&  \hskip -0.6cm S_a = \frac{1+{\rlap{v}/}}{2} \left[P_{1a}^{\prime \mu}\gamma_\mu\gamma_5-P_{0a}^*\right]   \nn \\
&&  \hskip -0.6cm T_a^\mu=\frac{1+{\rlap{v}/}}{2} \Bigg\{
P^{\mu\nu}_{2a} \gamma_\nu \label{pos2}  \\
&&  \hskip  0.6cm -P_{1a\nu} \sqrt{3 \over 2} \gamma_5
\left[g^{\mu \nu}-{1 \over 3} \gamma^\nu (\gamma^\mu-v^\mu)
\right]
\Bigg\} \nn  \\
&&  \hskip -0.6cm X_a^\mu =\frac{1+{\rlap{v}/}}{2} \Bigg\{
P^{*\mu\nu}_{2a} \gamma_5 \gamma_\nu \nn  \\
&&  \hskip  0.6cm -P^{*\prime}_{1a\nu} \sqrt{3 \over 2}  \left[
g^{\mu \nu}-{1 \over 3} \gamma^\nu (\gamma^\mu-v^\mu) \right]
\Bigg\}   \nn   \\
&&  \hskip -0.6cm X_a^{\prime \mu\nu} =\frac{1+{\rlap{v}/}}{2}
\Bigg\{ P^{\mu\nu\sigma}_{3a} \gamma_\sigma
 -P^{*'\alpha\beta}_{2a} \sqrt{5 \over 3}
\gamma_5 \Bigg[ g^\mu_\alpha g^\nu_\beta
\nn  \\
&&  \hskip  0.6cm-{1 \over 5} \gamma_\alpha g^\nu_\beta
(\gamma^\mu-v^\mu) -{1 \over 5} \gamma_\beta g^\mu_\alpha
(\gamma^\nu-v^\nu) \Bigg] \Bigg\} \nn \eea with the various
operators annihilating mesons of four-velocity $v$ (conserved in
strong interactions) and   containing a factor $\sqrt{m_P}$. Light
pseudoscalars are introduced using
 $\displaystyle \xi=e^{i {\cal M} \over
f_\pi}$,with: \bea  \small {\cal M}= \left(\begin{array}{ccc}
\sqrt{\frac{1}{2}}\pi^0+\sqrt{\frac{1}{6}}\eta & \pi^+ & K^+\nonumber\\
\pi^- & -\sqrt{\frac{1}{2}}\pi^0+\sqrt{\frac{1}{6}}\eta & K^0\\
K^- & {\bar K}^0 &-\sqrt{\frac{2}{3}}\eta
\end{array}\right)\nn
\eea ($f_{\pi}=132 \; $ MeV).  At the leading order in the heavy
quark mass and light meson momentum expansion the decays  $F \to H
M$ $(F=H,S,T,X,X^\prime$ and $M$ a light pseudoscalar meson) can
be described by the Lagrangian interaction  terms (invariant under
chiral and
 heavy-quark spin-flavour transformations)
\cite{hqet_chir,casalbuoni}:
\bea && \hskip -0.6cm {\cal L}_H = \,  g \, Tr [{\bar H}_a H_b
\gamma_\mu \gamma_5 {\cal
A}_{ba}^\mu ] \nn \\
&& \hskip -0.6cm {\cal L}_S =\,  h \, Tr [{\bar H}_a S_b
\gamma_\mu \gamma_5 {\cal
A}_{ba}^\mu ]\, + \, h.c. \,,  \label{lag-hprimo} \\
&& \hskip -0.6cm {\cal L}_T =  {h^\prime \over
\Lambda_\chi}Tr[{\bar H}_a T^\mu_b (i D_\mu {\spur {\cal
A}}+i{\spur D} { \cal A}_\mu)_{ba} \gamma_5
] + h.c.   \nn  \\
&& \hskip -0.6cm {\cal L}_X =  {k^\prime \over
\Lambda_\chi}Tr[{\bar H}_a X^\mu_b (i D_\mu {\spur {\cal
A}}+i{\spur D} { \cal A}_\mu)_{ba} \gamma_5
] + h.c.  \nn \\
&& \hskip -0.6cm {\cal L}_{X^\prime} =  {1 \over
{\Lambda_\chi}^2}Tr[{\bar H}_a X^{\prime \mu \nu}_b [k_1 \{D_\mu,
D_\nu\} {\cal A}_\lambda  \nn \\
&& \hskip 0.4cm + k_2 (D_\mu D_\nu { \cal A}_\lambda + D_\nu
D_\lambda { \cal A}_\mu)]_{ba}  \gamma^\lambda \gamma_5] + h.c.
\nn \eea
where  $D_{\mu ba}
=-\delta_{ba}\partial_\mu+\frac{1}{2}\left(\xi^\dagger\partial_\mu
\xi +\xi\partial_\mu \xi^\dagger\right)_{ba}$, ${\cal A}_{\mu
ba}=\frac{i}{2}\left(\xi^\dagger\partial_\mu \xi-\xi
\partial_\mu \xi^\dagger\right)_{ba}$.
$\Lambda_\chi$ is  the chiral symmetry-breaking scale; we  use
$\Lambda_\chi = 1 \, $ GeV. ${\cal L}_S$ and ${\cal L}_T$ describe
transitions of positive parity heavy mesons with the emission of
light pseudoscalar mesons in $s-$ and $d-$ wave, respectively, $g,
h$ and $h^\prime$ representing  effective coupling constants.  On
the other hand, ${\cal L}_X$ and ${\cal L}_{X^\prime}$  describe
the transitions of higher mass mesons of negative parity with the
emission of light pseudoscalar mesons in $p-$ and $f-$ wave with
coupling constants $k^\prime$, $k_1$ and $k_2$. We only consider
these terms: the light meson momenta involved in the
$D_{sJ}(2860)$ decays are $q_K=0.59 $ GeV for $D^*K$ final state
  and $q_K=0.7 $ GeV for $DK$ final state, so  it is possible that other terms
  in the light-meson momentum expansion, involving other structures and couplings,
  should be taken into account
in the interaction Lagrangian. At present these terms are unknown.

At the same  order in the expansion in the light meson momentum,
the structure of the Lagrangian terms for radial excitations of
the $H$, $S$ and $T$ doublets does not change,  since it  is only
dictated by the spin-flavour and chiral symmetries, but  the
coupling constants $g, h$ and $h^\prime$ have to be  substituted
by $\tilde g, \tilde h$ and $\tilde h^\prime$. The advantage of
this formulation is that meson transitions  into final states
obtained by $SU(3)$ and heavy quark spin rotations can be related
in a straightforward way.

Let us start with $D_{sJ}(2860)$. A new $c \bar s$ meson decaying
to $DK$ can be either  the $J^P=1^-$ state of the  $ s_\ell^P={3
\over 2}^-$ doublet,  or the $J^P=3^-$ state of the  $ s_\ell^P={5
\over 2}^-$ one, in both cases with  lowest radial quantum number.
Otherwise $D_{sJ}(2860)$ could be a radial excitation of already
observed $c\bar s$ mesons: the first radial excitation of $D_s^*$
($J^P=1^-$   $ s_\ell^P={1 \over 2}^-$) or of $D_{sJ}(2317)$
($J^P=0^+$   $ s_\ell^P={1 \over 2}^+$) or  of $D_{s2}^*(2573)$
($J^P=2^+$ $ s_\ell^P={3 \over 2}^+$).

As for $D_{sJ}(2710)$, two possibilities  can   be considered,
since the spin is known:
\begin{itemize}
\item $D_{sJ}(2710)$ belongs to the $ s_\ell^P={1 \over 2}^-$
doublet    and is the first radial excitation ($D_s^{* \prime}$);
\item $D_{sJ}(2710)$ is the low lying state with $ s_\ell^P={3
\over 2}^-$ ($D_{s1}^{* }$).
\end{itemize}

In \cite{Colangelo:2006rq} and \cite{Colangelo:2007ds} we
investigated  the decay modes of $D_{sJ}(2860)$ and $D_{sJ}(2710)$
according to the various possible assignments with the aim of
discriminating among them. The results are collected in Table
\ref{ratios}, where we report the ratios \bea R_1&=&
{\Gamma( D_{sJ}  \to D^*K) \over \Gamma( D_{sJ} \to DK) } \nn \\
R_2&=&{\Gamma( D_{sJ} \to D_s \eta) \over \Gamma( D_{sJ} \to DK) }
\eea (with $D^{(*)}K=D^{(*)+} K_S +D^{(*)0} K^+$) obtained for
various quantum number assignments to $D_{sJ} (2860)$ and
$D_{sJ}(2710)$ on the basis of
 eqs. (\ref{pos2}) and (\ref{lag-hprimo}). Also the decay
 to $D^*_s \eta$ is allowed, but it is  suppressed due to the limited phase space available.
 The ratios
 do not depend on the coupling constants,  but only on the quantum numbers.
%
\begin{table}[bt]
\caption{Predicted ratios
    $R_1$ and $R_2$ (see text for definitions)
 for the various assignment
 of quantum numbers to  $D_{sJ}(2860) $ and $D_{sJ}(2710)$.  }
    \label{ratios}
    \begin{center}
    \begin{tabular}{| c |  c | c |}\hline
 $D_{sJ}(2860) $  &$R_1$  &   $R_2$
\\ \hline
 $s_\ell^p={1\over 2}^-$, $J^P=1^-$,  $n=2$  &$1.23$& $0.27$ \\
$s_\ell^p={1\over 2}^+$, $J^P=0^+$, $n=2$   &$0$& $0.34$ \\
$s_\ell^p={3\over 2}^+$, $J^P=2^+$, $n=2$   &$0.63$& $0.19$\\
$s_\ell^p={3\over 2}^-$, $J^P=1^-$,   $n=1$   & $0.06$& $0.23$ \\
$s_\ell^p={5\over 2}^-$, $J^P=3^-$,   $n=1$   & $0.39$& $0.13$ \\
    \hline \hline $D_{sJ}(2710) $  &$R_1$  &   $R_2$
\\ \hline $s_\ell^p={1\over 2}^-$, $J^P=1^-$,  $n=2$ & $0.91 $ & $0.20 $   \\ \hline
  $s_\ell^p={3\over 2}^-$, $J^P=1^-$,   $n=1$ & $0.043 $ & $0.163 $
  \\ \hline
    \end{tabular}
  \end{center}
\end{table}
We first discuss  the entries in Table \ref{ratios} which concern
$D_{sJ}(2860)$.

The case $s_\ell^p={3\over 2}^-$, $J^P=1^-$, $n=1$ can  be
excluded since,   using the relevant  term  in (\ref{lag-hprimo})
and $k^\prime \simeq h^\prime\simeq 0.45\pm0.05$ (as  the
$h^\prime$   was determined in \cite{Colangelo:2005gb}),
 would give  $\Gamma(D_{sJ}\to DK)\simeq 1.5$ GeV, a result
 incompatible with the measured width.

In the  assignment  $s_\ell^p={1\over 2}^+$, $J^P=0^+$, $n=2$ the
decay to $D^*K$ is forbidden.  However, if $D_{sJ}(2860)$ is a
scalar radial excitation, it should have  a spin partner with
 $J^P=1^+$ ($s_\ell^p={1\over 2}^+$, $n=2$)  decaying to $D^* K$ with a small width,
  a rather easy signal to detect.
 For $n=1$ both  $D_{sJ}(2317)$ and $D_{sJ}(2460)$ are produced in charm continuum
 at  $e^+ e^-$ factories.
To explain the absence of the $D^*K$ in charm continuum events  at
mass around $2860$ MeV, one should  invoke some mechanism favoring
the production of the $0^+$ $n=2$ state and inhibiting the
production of $1^+$ $n=2$ state,
 a mechanism which discriminates the first radial excitation from the low lying state $n=1$.
 Such a mechanism is  difficult to imagine
 \cite{footnote1}.

Among the remaining possibilities, the assignment
$s_\ell^p={5\over 2}^-$, $J^P=3^-$, $n=1$ seems  the most likely
one.  In fact, in this case the small $DK$ width is due to the
 suppression related to the kaon momentum factor:
$\displaystyle \Gamma(D_{sJ}\to DK) \propto q_K^7$. The spin
partner would be $D_{s2}^{*}$, the $s_\ell^P={5 \over 2}^-$,
$J^P=2^-$ state,
 which can  decay to $D^* K$ and not to $DK$. It
would also  be narrow but  only  in the $m_Q\to \infty$ limit,
where the transition  $D_{s2}^{*}\to D^*K$ occurs in $f$-wave. As
an effect of $1/m_Q$ corrections this  decay can occur in
$p$-wave, so that   $D_{s2}^{*}$ could be  broader; therefore,  it
is not necessary to invoke a mechanism inhibiting the production
of this state with respect to $J^P=3^-$. If  $D_{sJ}(2860)$ has
$J^P=3^-$, it is not expected to  be produced
 in non leptonic $B$ decays such as
$B \to  D D_{sJ}(2860)$: the non leptonic amplitude in the
factorization approximation vanishes   since the vacuum matrix
element of the weak $V-A$ current with a spin three particle is
zero. Therefore,  the quantum number assignment can be confirmed
by studies of $D_{sJ}$ production in $B$ transitions. Actually, in
the Dalitz plot analysis of $B^+ \to \bar D^0 D^0 K^+$ Belle
Collaboration \cite{Brodzicka:2007aa} has found  no signal of
$D_{sJ}(2860)$.

The conclusion    is that $D_{sJ}(2860)$ is likely  a $J^P=3^-$
state,
 a  predicted high mass and relatively narrow $c \bar s$ state
 \cite{Colangelo:2000jq}. Its non-strange partner  $D_3$, if the mass splitting
$M_{D_{sJ}(2860)}-M_{D_3}$ is of the order of the strange quark
mass, is  also expected to be narrow: $\Gamma(D_{3}^+\to D^0 \pi^+
)\simeq 37$ MeV. It  can  be produced in semileptonic as well as
in non leptonic $B$ decays, such as $\bar B^0 \to D_3^+ \ell^-
\bar \nu_\ell$ and $\bar B^0 \to D_3^+ \pi^-$
\cite{Colangelo:2000jq}: its observation could  be used to assign
the proper quantum numbers to the  resonance $D_{sJ}(2860)$ found
by BaBar. Before considering the $D^*K$ mode, let us look at
 $D_{sJ}(2710)$. As  Table \ref{ratios} shows,
$R_1$ is very different if $D_{sJ}(2710)$ is $D_s^{*\prime}$ or
$D_{s1}^*$:
  the  $D^*K$  mode is the main  signal to be investigated in order to
distinguish between the two  possible assignments. From the
computed widths, assuming that  $\Gamma(D_{sJ}(2710))$ is
saturated by modes with a heavy meson and a light pseudoscalar
meson in the final state, we can determine the couplings $\tilde
g$ and $k^\prime$ governing the decays in the two cases.
Identifying $D_{sJ}(2710)$ with $D_s^{*\prime}$ we obtain: \be
\tilde g= 0.26 \pm 0.05 \,\,, \label{gtilde}\ee while if
 $D_{sJ}(2710)$ is $D_{s1}^*$ we get \be k^\prime=0.14 \pm
0.03 \,\,. \label{kprimo} \ee These  values are similar to those
obtained for analogous couplings  appearing in the effective heavy
quark chiral Lagrangians  \cite{altri,Colangelo:1995ph}.

The results for $\tilde g$ and $k^\prime$ can provide information
about the spin partner of $D_{sJ}(2710)$, i.e. the state belonging
to the same  $s_\ell^P$ doublet   from which $D_{sJ}(2710)$
differs only for  the total spin. The partner of $D_s^{*\prime}$
($s_\ell^P={1 \over 2}^-$) has $J^P=0^-$; it is denoted
$D_s^\prime$,   the first radial excitation of $D_s$, while the
partner of $D_{s1}^*$ ($s_\ell^P={3 \over 2}^-$) is the state
$D_{s2}^*$ with $J^P=2^-$. In both cases,  the decay modes to $
D^{*0} K^+$, $ D^{*+} K^0_{S(L)}$, $ D^{*}_s \eta$, are permitted.
In the heavy quark limit, these partners are degenerate.  Using
the obtained values for $\tilde g$ and $k^\prime$,  we get: \be
\Gamma(D_s^\prime)= (70 \pm 30) \,\, {\rm MeV} \,\,, \ee and
\be\Gamma(D_{s2}^*)= (12 \pm 5) \,\, {\rm MeV} \,\,, \ee  so that
in the two assignments the spin partners differ for their  total
width.

Along the same lines, one can   study the charmed mesons with the
same quantum numbers as $D_{sJ}(2700)$, but with a different light
quark flavour. These states are a charged charmed meson and a
neutral one, denoted as  $D_J^+$ and $D_J^0$, respectively. They
have not been observed yet,  so that their masses are unknown. We
assume such masses to be $2600 \pm 50$ MeV by the reasonable
assumptioncriterion that $D_{sJ}(2700)$ is heavier by an amount of
the size of the strange quark mass.

Allowed decay modes for $D_J^+(2600)$ are: $D_J^+  \to D^0 \pi^+$,
$D^+ \pi^0$,  $D_s {\bar K}_{S(L)}^0$, $D^+ \eta$,  and  $D_J^+
\to D^{*0}\pi^+$, $D^{*+}\pi^0$, $D^{*+} \eta$, while for
$D_{J}^0$ they are:
 $D_J^0  \to D^+ \pi^-$, $D^0 \pi^0$, $D_s  K^-$, $D^0 \eta$ and $D_J^0
\to D^{*0}\pi^0$, $D^{*+}\pi^-$, $D^{*0}\eta$;
 the corresponding widths  depend on  the
possible identification of  $D_{J}^{+(0)}$. The states  having
$s_\ell=\displaystyle{1 \over 2}^-$ are  denoted as
$D^{*\prime+(0)}$ and are radial excitations, while the states
having $s_\ell=\displaystyle{3 \over 2}^-$  are  denoted as
$D^{*+(0)}_1$.

Using  the effective coupling constants ${\tilde g}$ and
$k^\prime$  in (\ref{gtilde}), (\ref{kprimo}), we obtain:

\be \Gamma(D^{*\prime+(0)})=(128 \pm 61) \,\,{\rm MeV} \ee \be
 \Gamma(D_1^{*+(0)})= (85 \pm 46)\,\,{\rm MeV}
 \label{totwid-ud-part}\ee
so that  the $c \bar q$  partners have widths which are different
in the case of
 the two  assignments.  The mesons are not very broad, hence
it should be possible to observe them.

We conclude this discussion mentioning that a new experimental
analysis of $DK$ and $D^*K$ final states has been performed by
BaBar Collaboration \cite{Aubert:2009di}.

As it emerged above, the $D^*K$ mode plays an important role in
this context.  BaBar has observed both $D_{sJ}(2710)$ and
$D_{sJ}(2860)$ decaying to $DK$ and $D^*K$ final states, hence the
states should have natural parity $J^P=1^-, \, 2^+, \, 3^-,
\dots$. The assignment $J^P=0^+$ for $D_{sJ}(2860)$ is excluded.
More information comes from the measurement of the ratios
\cite{Aubert:2009di}: \bea {BR(D_{sJ}(2710) \to D^*K) \over
BR(D_{sJ}(2710) \to
DK)}=&& \hskip -0.5cm 0.91 \pm 0.13_{stat} \pm 0.12_{syst} \nn \\
{BR(D_{sJ}(2860) \to D^*K) \over BR(D_{sJ}(2860) \to DK)}=&&
\hskip -0.5cm 1.10 \pm 0.15_{stat} \pm 0.19_{syst} \nn \,\,.\eea
Comparing these data with the predictions in Table \ref{ratios},
one
concludes that \\
\begin{itemize}
\item $D_{sJ}(2710)$ is most likely $D_s^{*\prime}$, i.e. the
first radial excitation of $D_s^*(2112)$; \item the ratio
involving $D_{sJ}(2860)$ decays differs from the prediction at the
level of three standard deviations. The identification of this
state still requires further theoretical and experimental study
 both aiming at estimating the accuracy of the
predictions in table \ref{ratios} and of the experiments
\cite{footnote2}.
\end{itemize}

The last remark concerns the BaBar observation
 of
 another $c{\bar s}$ broad structure, with \cite{Aubert:2009di}:
 \bea
M&=& 3044 \pm 8_{stat}
(^{+30}_{-5})_{syst} \,\,\, {\rm MeV} \nn \\
\Gamma &=& 239 \pm 35_{stat} (^{+46}_{-42})_{syst} \,\,\, {\rm
MeV} \,\,.\nn \eea

Studies of angular distributions for this state have not been
attempted at present, due to the limited statistics. The
theoretical analysis of this state will be reported elsewhere.

\section{Symmetry breaking terms}

Heavy quark symmetries, holding in the infinite heavy quark mass
limit, are broken by terms
 which are suppressed by increasing powers of $m_Q^{-1}$
\cite{Falk:1995th}. Mass degeneracy between the members of the
meson doublets is broken by the  terms: \bea {\cal L}_{1/m_{Q}}={1
\over 2 m_{Q}} &\cdot&\big\{ \lambda_H Tr [{\bar H}_{a}
\sigma^{\mu \nu} H_{a} \sigma_{\mu \nu}]\nn \\ &&- \lambda_S
Tr [{\bar S}_{a} \sigma^{\mu \nu} S_{a} \sigma_{\mu \nu}]\nn \\
&&+ \lambda_T Tr [{\bar T}^\alpha_{a} \sigma^{\mu \nu}
T^\alpha_{a} \sigma_{\mu \nu}] \big\} \,\,\,\, \label{mass-viol}
\eea where the constants $\lambda_H$, $\lambda_S$ and $\lambda_T$
are related  to the hyperfine mass splittings: \bea \lambda_H &=&
{1 \over 8} \left( M_{P^*}^2-M_P^2 \right) \nn \\\lambda_S &=& {1
\over 8} \left( M_{P^\prime_1}^2-M_{P_0^*}^2 \right)  \\\lambda_T
&=& {3 \over 8} \left( M_{P^*_2}^2-M_{P_1}^2 \right) \,\,\, . \nn
\label{lambdas}\eea

Other two  effects related to spin symmetry-breaking terms concern
the possibility that  the members of the $s_\ell={3\over 2}^+$
doublet can  decay in S wave into the lowest lying heavy mesons
and pseudoscalars, and that a mixing may be induced between the
two $1^+$ states belonging to the two positive parity doublets
with different $s_\ell$. The corresponding terms in the effective
Lagrangian are: \bea {\cal L}_{D_1}={f \over 2m_Q \Lambda_\chi}&&
\hskip -0.2cm  Tr [{\bar H}_{a} \sigma^{\mu \nu} T^\alpha_{b}
\sigma_{\mu \nu} \gamma^\theta \gamma_5 \nn \\&& \hskip -0.2cm (i
D_\alpha {\cal A}_\theta+ iD_\theta {\cal A}_\alpha)_{ba}] + h.c.
\, \label{ld1} \eea \be {\cal L}_{mix}={b_1 \over 2m_Q} Tr[{\bar
S}_{a} \sigma^{\mu \nu}T_{\mu a} \sigma_{ \nu \alpha}v^\alpha] \,
+ h.c. \label{lmix}\ee
 Notice that ${\cal L}_{D_1}$ describes both $S$ and $D$ wave decays.
The mixing angle between the two $1^+$ states: \bea
\ket{P_1^{phys}}&=&\cos \theta
\ket{P_1}+ \sin{\theta} \ket{P_1^\prime} \label{d1phys} \\
\ket{P_1^{\prime phys}}&=&-\sin \theta \ket{P_1}+ \cos{\theta}
\ket{P_1^\prime} \label{d1primephys} \eea
 can be related to the coupling constant $b_1$ and to the mass splitting:
\be \tan \theta={\sqrt{\delta^2+\delta_g^2}-\delta \over \delta_g}
\label{mix-angle} \ee
 where
$\delta=\displaystyle{\Delta_T -\Delta_S \over 2}$,
$\delta_g=-\displaystyle{\sqrt{2 \over 3}{b_1 \over m_Q}}$ and the
mass parameters $\Delta_S$ and $\Delta_T$ which represent the mass
splittings between   positive and negative parity doublets. They
can be expressed
 in terms of the spin-averaged masses:
 $\Delta_S= \overline M_S - \overline M_H$ and  $\Delta_T= \overline M_T - \overline M_H$
with
\bea {\overline M}_H& =& {3 M_{P^*}+M_P  \over 4} \nn \\
{\overline M}_S &=& {3 M_{P^\prime_1}+M_{P_0^*} \over 4}  \\
{\overline M}_T &=& {5 M_{P^*_2}+3M_{P_1} \over 8} \,\, .\nn
\label{lambdapar}\eea
 The parameters  in the various terms of the
effective Lagrangian are universal and their determination  is
important in the definition of the effective theory and
 in the applications to the hadron phenomenology. Data recently collected on
 charmed and charmed-strange mesons, together with information on previously known
 positive parity charmed states, allow us to determine some of
 them.
 We identify  $D_{sJ}(2317)$ and
$D_{sJ}(2460)$ with the members of the
$J^P_{s_\ell}(0^+,1^+)_{1/2}$ doublet and, using with the masses
of the other charmed states reported in the PDG \cite{PDG}, we
obtain the values of $\lambda_H$, $\lambda_S$ and $\lambda_T$
reported in Table \ref{lambdaHST} \cite{Colangelo:2005gb}.
\begin{table}[h]
    \caption{ $\lambda_{i}$ parameters    obtained using data in PDG \cite{PDG}.
   The  spin-averaged masses
  for the various doublets and  the mass splittings $\Delta_{S}$
and $\Delta_{T}$ are also reported.}
    \label{lambdaHST}
    \begin{center}
    \begin{tabular}{|l|c|c|}
      \hline
  & $c{\bar q}$  & $c{\bar s}$ \\ \hline $\lambda_H$ & $ (261.1 \pm 0.7\,\, {\rm MeV})^2$ &
 $ (270.8 \pm 0.8 \,\, {\rm MeV})^2$
  \\
$\lambda_S$ & $ (265 \pm 57 \,\, {\rm MeV})^2$ &$(291 \pm 2 \,\,
{\rm MeV})^2$
 \\
$\lambda_T$ & $(259\pm 10 \,\, {\rm MeV})^2$ &$ (266 \pm 6\,\,
{\rm MeV})^2$ \\  \hline
$\overline M_{H}$ & $1974.8\pm 0.4 \, {\rm MeV}$ & $2076.1\pm 0.5 \, {\rm MeV}$  \\
$\overline M_{S}$ & $2397\pm 28 \, {\rm MeV}$ & $2424\pm 1 \, {\rm MeV}$ \\
$\overline M_{T}$ & $2445.1\pm 1.4 \, {\rm MeV}$ & $2558 \pm 1 \,
{\rm MeV}$\\ \hline
$ \Delta_{S}$ & $422\pm 28 \, {\rm MeV}$ & $348\pm 1 \, {\rm MeV}$  \\
$ \Delta_{T}$ & $470.3\pm 1.5 \, {\rm MeV}$ & $482\pm 1 \, {\rm MeV}$  \\
 \hline
    \end{tabular}
  \end{center}
\end{table}

In the above determinations we have neglected the mixing  angle
between the two $1^+$ states $D_1$ and $D_1^\prime$. Considering,
instead,
 the  result  $\theta_c=-0.10 \pm 0.03 \pm 0.02\pm 0.02 \,\,rad$ \cite{Abe:2003zm}
 and  using $\Delta_T$ and $\Delta_S$ in Table \ref{lambdaHST}
 together with eq. (\ref{mix-angle}) and $m_{c}=1.35 \,$ GeV,
we can compute the coupling $b_1$ in (\ref{lmix}): \be
b_1=0.008\pm 0.006  \,\, {\rm GeV}^2 \,\,, \ee  therefore
compatible with zero.

\begin{figure}[htb]
\centering
\includegraphics*[width=65mm]{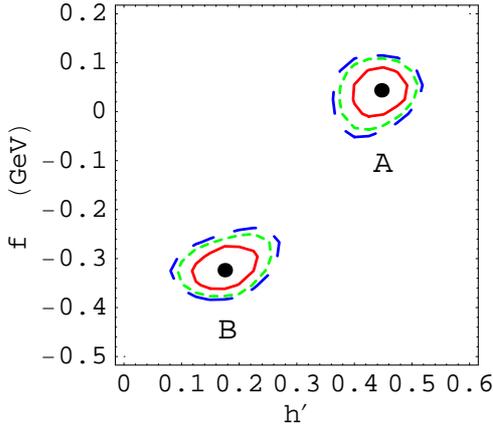}\\
\caption{Regions in the $(h^\prime,f)$ plane
 constrained by  the  widths of $D_2^{*0}$ and $D_1^0$. Only the region A is also compatible with
  the  constraints on the parameter $R$ in eq.(\ref{Rpar}).}
\label{regions}
\end{figure}

To determine the couplings $h^\prime$  and $f$ in eqs.
(\ref{lag-hprimo})-(\ref{ld1})   we consider
 the widths of the two members of the  $c{\bar q}$   $s_\ell^P={3 \over 2}^+$
doublet,  $D_1$ and $D_2^*$ together with  recent
 results  from Belle Collaboration \cite{Abe:2003zm}:
\bea \Gamma(D_2^{*0})&=& 45.6 \pm 4.4 \pm 6.5 \pm 1.6
\,\,\, {\rm MeV} \nn \\
\Gamma(D_1^{0}) &=& 23.7 \pm 2.7 \pm 0.2 \pm 4.0 \,\,\, {\rm MeV}
\label{exp-gamma} \,\,. \eea
In the plane $(h^\prime,f)$ four regions are allowed by  data
which, due to symmetry $(h^\prime, f) \to (-h^\prime, -f)$,
 reduce to the two inequivalent regions depicted in fig. \ref{regions}.

 A
further constraint is the  Belle
 measurement of the helicity angle distribution in the
decay $D_{s1}(2536) \to D^{*+} K_S^0$,  with the determination of
the ratio
 \be R= \displaystyle{\Gamma_s \over
\Gamma_s+\Gamma_d } \label{Rpar}\,\,\, ,\ee $\Gamma_{s,d}$  being
the $s$ and $d$ wave partial widths, respectively
\cite{Abe:2005xj}:
 $0.277 \le R \le 0.955$
(a measurement of the ratio $R$ versus the phase difference
between $s$ and $d$  was obtained by CLEO Collaboration for
non-strange mesons  \cite{Avery:1994yc}). Although the range of
$R$ is  wide,  it allows to exclude  the  region $B$ in
fig.\ref{regions},
 leaving only the region $A$  that can be represented as
\be h^\prime= 0.45 \pm 0.05  \hspace*{1cm}  f=0.044 \pm 0.044
\,\,{ \rm GeV}\label{results} \,\,.\ee The coupling constant $f$
is    compatible with zero,  hence the contribution of the
lagrangian term (\ref{ld1})  is small. Since also the coupling
$b_{1}$ turns out to be small, the two $1^{+}$ states
corresponding to the $s_{\ell}^{P}={1 \over 2}^{+},{3 \over 2}^{+}
$ practically coincide with the physical states. For
 the   width of $D_{s1}(2536)$ we predict
\be \Gamma(D_{s1}(2536))=2.5 \pm 1.6 \,\,\,{\rm MeV}
\label{2536wid} \ee
 compatible with the present  bound:
$\Gamma(D_{s1}(2536))<2.3$ MeV \cite{PDG}.

%

\section{Conclusions}

Our knowledge of charm spectroscopy has greatly improved in recent
years, but there are results which challenge   our understanding
of some aspects of Quantum Chromodynamics. The question whether
the many newly observed states are conventional or exotic ones has
been put forward in several cases, stimulating numerous
investigations. We have discussed the most recently observed
$c{\bar s}$ mesons, adopting a classification in terms of doublets
provided in the heavy quark limit.

\begin{table*}[tb]
\centering \caption{ $c \bar s$ states organized according to
$s_\ell^P$ and $J^P$. The mass of  known mesons    is indicated.
States in bold face have been placed in the table according  to
the interpretation supported in this paper.}
    {\begin{tabular}{|c |c c c c c|}\hline
 $s_\ell^P  $  &${1\over 2}^-$  &   ${1\over 2}^+$ &  ${3\over
2}^+$  & ${3\over 2}^-$ &
  ${5\over 2}^-$ \\ \hline $n=1$ & &&&& \\
      \hline
  $J^P=s_\ell^P-{1\over 2}$&$D_s (1965) \,\, (0^-)  $  & $D_{sJ}(2317) \,\, (0^+)$ &
$D_{s1}(2536) \,\, (1^+) $ & $(1^-)$& $(2^-)$\\
 $J^P=s_\ell^P+{1\over 2}$&$D_s^*(2112) \,\, (1^-)$ & $D_{sJ}(2460)   \,\, (1^+)$
  & $D_{s2}^*(2573) \,\,
  (2^+)  $ &  $(2^-)$&  $ {\bf D_{sJ}(2860)} \,\, (3^-)$\\
  \hline
  $n=2$ & &&&& \\
      \hline
  $J^P=s_\ell^P-{1\over 2}$&  $(0^-)  $& $(0^+)  $ & $(1^+)  $
 &  $(1^-)  $& $(2^-)  $ \\
 $J^P=s_\ell^P+{1\over 2}$& ${\bf D_{sJ}(2710)} \,\, (1^-)$ &  $(1^+)  $ & $(2^+)  $   & $(2^-)  $& $(3^-)  $\\
  \hline
   \end{tabular}} \label{schema}
\end{table*}

Our conclusion is represented by table \ref{schema}, where we
propose identification of  the states discussed above. In the
table, $D_{sJ}(2317)$ and $D_{sJ}(2460)$ are identified as the two
members of the $J^P_{s_\ell}=(0^+,1^+)_{1/2}$ doublet. As for the
other two states,  it is likely that the interpretation of
$D_{sJ}(2710)$ as the first radial excitation of $D_s^*$ is
correct. The identification of $D_{sJ}(2860)$  is still under
debate.

\vspace*{1.5cm} \noindent {\bf Acknowledgments} \\

\noindent I thank the organizers of CHARM09 for inviting me to
this interesting meeting.
 I also thank P. Colangelo, R. Ferrandes, S. Nicotri, A. Ozpineci and M. Rizzi
 for collaboration on the topics discussed in this paper.
 This work  was supported in part by the EU
contract No. MRTN-CT-2006-035482, "FLAVIAnet".


\begin{thebibliography}{99} 

\bibitem{reviews}
For reviews see: P.~Colangelo, F.~De Fazio and R.~Ferrandes,
  Mod.\ Phys.\ Lett.\  A { 19} (2004) 2083;
  E.~S.~Swanson,
  Phys.\ Rept.\  { 429} (2006) 243;
P.~Colangelo, F.~De Fazio, R.~Ferrandes and S.~Nicotri,
  Prog.\ Theor.\ Phys.\ Suppl.\  { 168} (2007) 202.


\bibitem{HQET} For  reviews see: M.~Neubert,
  Phys.\ Rept.\  { 245} (1994) 259;
  F.~De Fazio,
 in {\it At the
Frontier of Particle Physics/Handbook of QCD}, ed. by M. Shifman
(World Scientific, Singapore, 2001),  page  1671,
arXiv:hep-ph/0010007.



\bibitem{PDG}
C.~Amsler {\it et al.}  [Particle Data Group],
  Phys.\ Lett.\  B { 667} (2008) 1 and http://pdg.lbl.gov.

\bibitem{BaBar2317}
B.~Aubert {\it et al.}  [BABAR Collaboration],
 Phys. Rev. Lett.  { 90} (2003) 242001.

\bibitem{Belle continuo}
Y.~Mikami {\it et al.} [Belle Collaboration], Phys. Rev. Lett. {
92} (2004) 012002.

\bibitem{CLEO} D.~Besson {\it et al.}  [CLEO Collaboration],
Phys.\ Rev.\ D { 68} (2003) 032002.

\bibitem{FOCUS2317}
L.~Benussi  [FOCUS Collaboration],
  Int.\ J.\ Mod.\ Phys.\  A { 20} (2005) 549.


\bibitem{Godfrey:2003kg}
S.~Godfrey,
  Phys.\ Lett.\ B {568}  (2003) 254.

\bibitem{Colangelo:2003vg}
  P.~Colangelo and F.~De Fazio,
  Phys.\ Lett.\  B { 570} (2003) 180.

\bibitem{Colangelo:2005hv}
  P.~Colangelo, F.~De Fazio and A.~Ozpineci,
  Phys.\ Rev.\  D { 72} (2005) 074004.

\bibitem{altri}
   V.~M.~Belyaev, V.~M.~Braun, A.~Khodjamirian and R.~Ruckl,
Phys.\ Rev.\ D { 51}  (1995) 6177, and references therein.

\bibitem{Colangelo:2000dp}
For a review see: P.~Colangelo and A.~Khodjamirian, in {\it At the
Frontier of Particle Physics/Handbook of QCD}, ed. by M. Shifman
(World Scientific, Singapore, 2001),  page 1495,
arXiv:hep-ph/0010175.

\bibitem{Ball:2002ps}
P.~Ball, V.~M.~Braun and N.~Kivel,
Nucl.\ Phys.\ B { 649}  (2003) 263.

\bibitem{Colangelo:1995ph}
P.~Colangelo, F.~De Fazio and G.~Nardulli,
  Phys.\ Lett.\  B {\bf 334} (1994) 175;
P.~Colangelo, G.~Nardulli, A.~Deandrea, N.~Di Bartolomeo, R.~Gatto
and F.~Feruglio,
  Phys. Lett.  B 339 (1994) 151;
  P.~Colangelo, F.~De Fazio, G.~Nardulli,
  N.~Di Bartolomeo and R.~Gatto,
  Phys. Rev.  D 52 (1995) 6422;
  P.~Colangelo and F.~De Fazio,
  Eur. Phys. J.  C 4 (1998) 503;
P.~Colangelo and F.~De Fazio,
  Phys.\ Lett.\  B {\bf 532} (2002) 193.


\bibitem{Bardeen}
W.~A.~Bardeen, E.~J.~Eichten and C.~T.~Hill,
Phys.\ Rev.\ D { 68}  (2003) 054024.

\bibitem{Aubert:2006mh}
  B.Aubert,
  Phys. Rev. Lett.  97 (2006) 222001.

\bibitem{Brodzicka:2007aa}
  J.~Brodzicka {\it et al.},
  Phys. Rev. Lett.  100 (2008) 092001.

\bibitem{Gasser:1984gg}
  J.~Gasser and H.~Leutwyler,
  Nucl. Phys.  B 250 (1985) 465.


\bibitem{hqet_chir}
M.B. Wise, Phys. Rev. D 45   (1992) R2188; G. Burdman and J.F.
Donoghue, Phys. Lett. B 280  (1992) 287; P. Cho, Phys. Lett. B 285
(1992)  145; H.-Y. Cheng {\it et al.,}  Phys. Rev.  D 46
 (1992) 1148.

\bibitem{casalbuoni}
R.~Casalbuoni, A.~Deandrea, N.~Di Bartolomeo, R.~Gatto,
F.~Feruglio and G.~Nardulli,
 Phys. Lett.  B 299
(1993) 139.


\bibitem{Colangelo:2006rq}
  P.~Colangelo, F.~De Fazio and S.~Nicotri,
  Phys. Lett.  B 642 (2006) 48.

\bibitem{Colangelo:2007ds}
  P.~Colangelo, F.~De Fazio, S.~Nicotri and M.~Rizzi,
  Phys.\ Rev.\  D { 77} (2008) 014012.

\bibitem{Colangelo:2005gb}
   P.~Colangelo, F.~De Fazio and R.~Ferrandes,
  Phys. Lett.  B 634 (2006) 235.

\bibitem{footnote1}
The interpretation of $D_{sJ}(2860)$ as the first radial
excitation of $D_{sJ}(2317)$  has been proposed in
  E.~van Beveren and G.~Rupp,
  Phys. Rev. Lett.   97 (2006) 202001;
  F.~Close {\it et al.},
  Phys. Lett. B 647 (2007) 159.

\bibitem{Colangelo:2000jq}
  P.~Colangelo, F.~De Fazio and G.~Nardulli,
  Phys. Lett. B478 (2000) 408.

\bibitem{Aubert:2009di}
  B.~Aubert  [The BABAR Collaboration],
  arXiv:0908.0806 [hep-ex].

\bibitem{footnote2}
 The possibility that  $D_{sJ}(2860)$ is a mixing of
$J^P=0^+$ and $J^P=2^+$ components is suggested in
 E.~van Beveren and G.~Rupp,
  arXiv:0908.1142 [hep-ph].
  .

\bibitem{Falk:1995th}
  A.~F.~Falk and T.~Mehen,
  Phys.\ Rev.\ D {53} (1996) 231.

\bibitem{Abe:2003zm}
  K.~Abe {\it et al.}  [Belle Collaboration],
  Phys.\ Rev.\ D { 69} (2004) 112002.


\bibitem{Abe:2005xj}
  K.~Abe {\it et al.}  [Belle Collaboration],
  arXiv:hep-ex/0507030.

\bibitem{Avery:1994yc}
  P.~Avery {\it et al.}  [CLEO Collaboration],
  Phys.\ Lett.\ B { 331} (1994) 236
  [Erratum-ibid.\ B { 342} (1995) 453].


\end{thebibliography}
\end{document}